\documentclass[aps,prl,twocolumn,floatfix,superscriptaddress]{revtex4}
\usepackage[]{graphicx}
\usepackage{dcolumn}
\usepackage{bm}
\usepackage{amsmath, amssymb}
\usepackage{subfigure}
\usepackage{color}

\begin{document}
\title{Quantum differential ghost microscopy}

\author{E. Losero}
\affiliation{Istituto Nazionale di Ricerca Metrologica (INRIM), Strada delle Cacce 91, 10135 Torino, Italy}
\affiliation{Politecnico di Torino, C.so Duca degli Abruzzi 24, 10129 Torino, Italy}

\author{I. Ruo-Berchera}
\author{A. Meda}
\author{A. Avella}
\affiliation{Istituto Nazionale di Ricerca Metrologica (INRIM), Strada delle Cacce 91, 10135 Torino, Italy}

\author{O. Sambataro}
\affiliation{Universit\`a degli Studi di Torino, Dipartimento di Fisica, Via P. Giuria 1, 10125 Torino, Italy}

\author{M. Genovese}
\affiliation{Istituto Nazionale di Ricerca Metrologica (INRIM), Strada delle Cacce 91, 10135 Torino, Italy}
\affiliation{INFN, sezione di Torino, via P. Giuria 1, 10125 Torino, Italy}

\begin{abstract}
Quantum correlations become formidable tools for beating classical capacities of measurement. Preserving these advantages in practical systems, where experimental imperfections are unavoidable, is a challenge of the utmost importance.
Here we propose and realize a novel quantum ghost imaging protocol stemming from the differential ghost imaging, a scheme elaborated so far in the limit of bright thermal light, particularly suitable in the relevant case of faint or sparse objects. The extension towards the quantum regime represents an important step as quantum correlations allow low brightness imaging, desirable for reducing the absorption dose. Furthermore, we optimize the protocol in terms of signal-to-noise ratio, to compensate for the detrimental effects of detection noise and losses.
We perform the experiment using SPDC light in a microscope configuration. The image is reconstructed exploiting non-classical intensity correlation  in the low photon flux regime, rather than photon pairs detection coincidences. On the one side, we validate the theoretical model and on the other we show the applicability of this technique by imaging biological samples.

%\textbf{Keywords:} imaging, ghost imaging, microscopy, parametric down conversion
\end{abstract}

\maketitle

\section*{Introduction}
%One of the first applications of quantum correlations has been the ghost imaging, a technique aimed to image an object detecting the photons that interacted with it without spatial reso

Ghost imaging (GI) was theoretically proposed in 1994 \cite{GI94} and experimentally demonstrated by Pittman et al. one year later \cite{GI95} by using quantum correlations generated by spontaneous parametric down conversion (SPDC). It was considered as the earliest quantum imaging technique \cite{Ruo19}, but soon it has been shown that also classical correlations, as present in split thermal beams, can be successfully exploited, although with smaller visibility \cite{thGI, thGI3, thGI4, thGI5, thGI6, hartmann15, Shapiro12}.
Two spatially correlated beams are used: one is addressed to the object to be imaged and then to a bucket detector, namely a detector without spatial resolution, while the other is addressed directly to a spatially resolving detector, without interacting with the object. Neither of the two beams separately contains information on the object absorption profile, nevertheless it can be retrieved exploiting the correlation between them. %Typically, reconstruction algorithms exploit the covariance between each pixel of the resolving detector and the bucket signal .
Since based on the evaluation of second-order momenta of the joint distribution, GI is not a single-shot technique, instead it requires the acquisition of several frames and the signal-to-noise ratio (SNR) scales with the square root of the number of frames \cite{review2017, genovese2016}.
%In the first ghost imaging demonstration quantum correlated beams generated by spontaneous parametric down conversion (SPDC) were employed, thus it was considered as the earliest quantum imaging technique, but soon it has been shown that also classical correlations, as present in split thermal beams, can be successfully exploited although with smaller visibility \cite{Ruo19, thGI, thGI3, thGI4, thGI5, thGI6, Shapiro12}.

Initially motivated by a fundamental debate on the boundary between classical and quantum resources, the attention toward GI is justified by the fact that it can be useful in various practical situations, as in the cases where environmental constraints do not allow placing fine optical system and pixeled detector behind the object \cite{GImagnet}, and because it is robust in presence of atmospheric turbulence or diffusive media on the object path \cite{turbulence, turbulence2, turbulence3}.

Quantum photon number correlations, a tool of the utmost relevance in quantum technologies \cite{Ruo19, genovese2016, review2017, qt1, qt2, qt3, qt4, qt5, qt6, qt7}, can add further advantages to GI as the possibility to probe the sample with a wavelength which differs from the one detected with spatial resolution  \cite{twocolorGI,Aspden15}, in case the first one is in a range where imaging system and spatial detection is technologically demanding. Moreover, quantum correlation allows retrieving ghost image at faint photon flux with a better SNR than by classical beams \cite{Aspden15,Brida11,Morris15}, where the stronger than classical correlation can be used also to reject external and detection noise \cite{qt2, Morris15,Kalashnikov14}.

%Note that reducing the photon dose can be relevant in several practical situation, as for example when using X-ray \cite{xray1} or investigating delicate samples [ref!!!].  investigation of photo-sensitive samples or delicate biological systems \cite{biodamage, biodamage2}.

%Ghost imaging (GI) technique was theoretically proposed in 1994 \cite{GI94} and experimentally demonstrated by Pittman et al. one year later \cite{GI95}. Two spatially correlated beams are used: one is addressed to the object to be imaged and then to a bucket detector, namely a detector without spatial resolution, while the other is addressed directly to a spatially resolving detector, without interacting with the object. Neither of the two beams separately contains information on the object absorption profile, nevertheless it can be retrieved exploiting the correlation between them. Typically, reconstruction algorithms exploit the covariance between the signal at each pixel of the resolving detector and the bucket signal \cite{genovese2016}. Since based on the evaluation of second-order momenta of the joint distribution, GI is not a single-shot technique, instead it requires the acquisition of several frames and the signal-to-noise ratio (SNR) scales with the square root of the number of frames.

Several variations of GI, aiming at increasing its applicability in realistic scenarios have been proposed, as for example back-scattering GI \cite{reflectiveGI}, computational GI \cite{computationalGI, computationalGI2}, compressive GI \cite{Katz09,Yu14}.

Among them, the so called "differential GI" (DGI) proposed in 2010 by Ferri et al. \cite{DGI} has received a large attention due to its relevant practical impact, addressing the problem of reconstructing small or faint objects in the field of view. In this situation the conventional GI typically  fails because it requires an unaffordable number of acquisitions in order to reach a sufficiently high SNR.
Proposed and experimentally realized only with bright thermal light \cite{DGI,turbulence2}, DGI does not involve changes in the typical GI optical setup, rather a more efficient use of the available data. In particular also the integrated signal in the reference channel, from the spatially resolving detector, is used in the data elaboration.

%It is important to investigate the possible extensions of the DGI approach to the quantum case, given its advantage in specific situations, for example in low brightness regime. In this work we address this point, developing a comprehensive theoretical model which takes into account non idealities such as channels inefficiencies and electronic noise of the detector, and performing the experiment.
In this work we study the performance of a DGI approach when exploiting quantum resources, motivated by their possible advantage in specific situations, for example in low brightness regime. We develop a comprehensive theoretical model of the SNR for both classical and quantum sources, which takes into account non idealities such as channels inefficiencies and electronic noise of the detector.
It comes out that the brightness of the source, namely the number of photons per spatio-temporal mode, has a fundamental role in determining DGI performances, in terms of SNR, together with the photon losses. In particular, for low-brightness sources, the improvement provided by DGI is dramatically affected by photon losses, disappearing or becoming worse than GI for loss probability larger than 50\%. These limitations of the DGI protocol cannot be derived from the classical description of Ref. \cite{DGI}, where correlated beams are treated as identical copies of the same classical stochastic process (an approximation that is suitable only in case of intense thermal beam).

However, we propose an optimized DGI protocol (ODGI) able to partially compensate for experimental imperfections, retrieving an advantage on GI for any value of losses and brightness. The only further requirement for its application is the characterization of channels efficiencies. This protocol can have positive impact where it is necessary to keep low the photon flux, as in X-ray GI \cite{xray1, xray2}. The optimization procedure stems from the one proposed and realized in the absorption estimation framework \cite{absorption, moreau}.

For demonstrating the performances of this method, we perform an experiment using SPDC light in the low brightness regime. However, rather than basing the GI reconstruction on temporal coincidences among photon pairs, as in the typical approach of almost all GI experiments with quantum light, here we exploit non-classical intensity correlation certified by the evaluation of a specific non-classicality parameter known as noise reduction factor (NRF) \cite{review2017,Brida2009,bondani2019,perina2018, checkova2016}.

We validate the model comparing its prediction with the SNR experimentally estimated for the GI, DGI and the ODGI protocols. Besides that, in view of real application, a biological object, a $(285 \mu \mathrm{m})^2$ wasp wing detail, is reconstructed with spatial resolution of 5 $\mu$m.

\section*{Theory}

\begin{figure}[tbp]
	\centering
	\includegraphics[ trim= 1.8cm 2cm 3cm 2cm, clip=true, width=0.49\textwidth, angle=0]{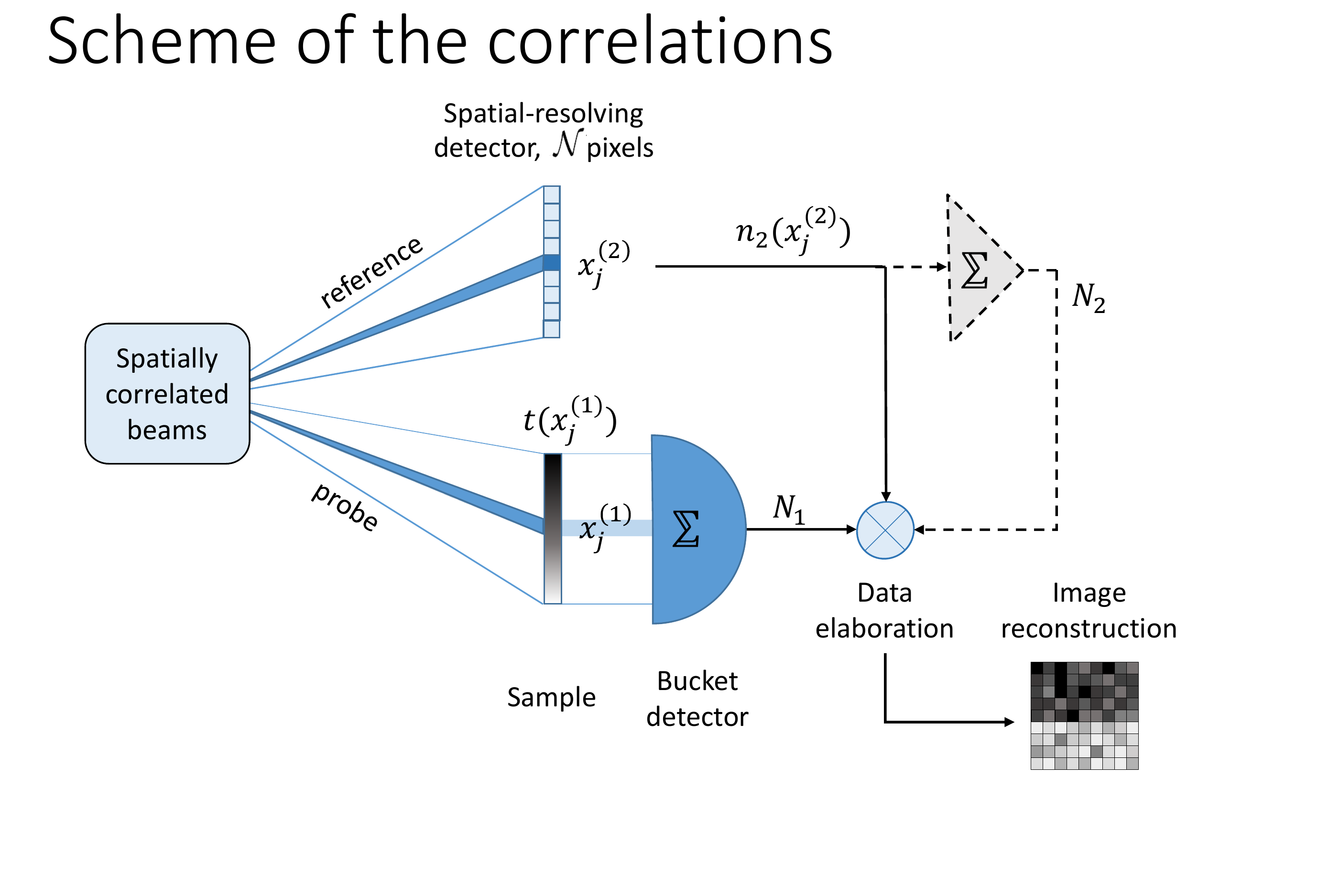}
	\caption{Scheme of GI and DGI protocols. Spatial correlations between two beams are exploited. The reference beam is detected by a spatial resolving detector, the probe beam, after interacting with the sample impinges on a detector without spatial resolution. Every pixel $x_j^{(2)}$ is in one-to-one correspondence with a resolution cell $x_j^{(1)}$ at the object plane. For the image reconstruction the number of photons detected by the bucket detector $N_1$ and the one detected by each pixel in the resolving detector, $n_2(x_j^{(2)})$, are used in the data elaboration. In DGI protocol (dashed path) also the integrated signal $N_2$ is exploited.}\label{corr_scheme}
\end{figure}

We consider either equally split multi-mode thermal beam ($th$) or multi-mode twin-beam ($tw$)  generated by SPDC. In both cases the two beams, used as probe and reference respectively, present spatial correlation \cite{review2017}.
In particular, referring to Fig. \ref{corr_scheme}, the spatial selection performed in the reference beam by the pixel in $x^{(2)}_j$ automatically identifies a small area where the correlated probe photons  are expected to impinge on the object  plane, centered in $x^{(1)}_j$. This area represents the spatial resolution of any GI scheme. This can be obtained, for example, if the point-to-point far-field correlations of SPDC are imaged at the detection plane for the reference beam, while at the object plane for the probe beam. Similar condition can be obtained by pairs of correlated spatial modes in split pseudothermal beams. We further assume that the pixel is larger than the coherence area, so that the resolution cells identified by two adjacent pixels do not overlap. Given this one-to-one correspondence between object plane and reference detection plane, hereinafter we will omit the suffixes 1 and 2, identifying $x^{(2)}_j$ and $x^{(1)}_j$.

Each beam of the twin-beam follows thermal statistics, therefore there is no difference between the thermal and twin beam case when  probe and reference beams are considered separately. In particular, the mean number of photons detected by the pixel in $x_j$ is:
\begin{equation}
\langle \hat{n}_2 (x_j) \rangle^{th} = \langle \hat{n}_2 (x_j) \rangle^{tw} = n_2
\end{equation}
where we have assumed a spatially uniform beam.
The corresponding number of photons passing through the resolution cell at the object plane and detected by the bucket is:
\begin{equation}
\langle \hat{n}_1 (x_j) \rangle^{th} = \langle \hat{n}_1 (x_j) \rangle^{tw} = n_2 \; t(x_j)
\end{equation}
with $t(x_j)$ the mean transmission profile of the object in the resolution cell.
Note that these expressions are obtained assuming the same loss level on the two channels in absence of the object $\eta_1=\eta_2=\eta$. This does not reduce the generality of our model since possible unbalancement between the channels can be included in the transmission profile of the object.
 When $M$  spatio-temporal modes are collected per pixel per frame, the associated variances can be written as \cite{review2017}:
\begin{equation}\label{var_n2}
\begin{split}
\langle \delta ^2 \hat{n}_2 (x_j) \rangle^{th} = \langle \delta ^2 \hat{n}_2 (x_j) \rangle^{tw} = \\ n_2 (1 + n_2/M) + \Delta_{el}^2
\end{split}
\end{equation}
\begin{equation}\label{var_n1}
\begin{split}
\langle \delta ^2 \hat{n}_1 (x_j) \rangle^{th} = \langle \delta ^2 \hat{n}_1 (x_j) \rangle^{tw} = \\
n_2  t(x_j) (1 + n_2  t(x_j) /M) + \Delta_{el}^2
\end{split}
\end{equation}
In Eq.s (\ref{var_n2}) and (\ref{var_n1}) we have taken into account the electronic noise of the detector, $\Delta_{el}$.
The crucial difference between thermal and twin beam case is in the covariance between two spatially correlated modes \cite{review2017}:
\begin{equation}\label{cov_Th}
\langle \delta \hat{n}_2 (x_j) \delta \hat{n}_1 (x_i) \rangle^{th} = \frac{t(x_i) n_2^2}{M} \delta_{i,j}
\end{equation}
\begin{equation}\label{cov_SPDC}
\langle \delta \hat{n}_2 (x_j) \delta \hat{n}_1 (x_i) \rangle^{tw} =t(x_i) \left( \frac{ n_2^2}{M} +  \eta n_2 \right) \delta_{i,j}
\end{equation}
It is immediately clear that the correlation is always higher for SPDC  light. The ratio of the two expressions is:
\begin{equation}\label{eq7}
\frac {\langle \delta \hat{n}_2 (x_j) \delta \hat{n}_1 (x_j) \rangle^{tw}}{\langle \delta \hat{n}_2 (x_j) \delta \hat{n}_1 (x_j) \rangle^{th}} = 1+\eta \frac{M}{n_2} %= 1 + \frac{1}{\mu}
\end{equation}
showing that the correlations, at the base of GI protocols, are significantly stronger for twin beam for small number of photons emitted per mode, $n_2 \ll \eta M$. It follows that, in this regime, only twin-beam light allows the object reconstruction in practical situations. In the opposite regime the performance of equally split thermal light and twin beam is asymptotically the same.

In the conventional GI protocol the transmission profile of the object is retrieved considering the covariance between each pixel of the reference channel $\hat{n}_2(x_{j})$ and the bucket detector on the other channel, $\hat{N}_1=\sum_{i=1}^{\mathcal{N}} \hat{n}_1 (x_i) $, as represented in Fig. \ref{corr_scheme}.  Here, $\mathcal{N}$ is the number of resolution cells at the object plane which corresponds to the number of pixels at the resolving detector. Indeed, from either Eq. (\ref{cov_Th}) or  Eq. (\ref{cov_SPDC}) it comes out that for both thermal and twin beam case, the reconstructed image is:
\begin{equation}
S_{\mathrm{GI}}(x_j)=\langle \delta \hat{N}_1\delta \hat{n}_2(x_{j})\rangle\propto t(x_j)
\end{equation}
In \cite{DGI}, Ferri et al. propose the DGI protocol, where the bucket detector signal $\hat{N}_1$ is replaced by $\hat{N}_{\mathrm{\mathrm{DGI}}} = \hat{N}_1 - \frac{\langle \hat{N}_1 \rangle}{\langle \hat{N}_2 \rangle} \hat{N}_2$, with  $\hat{N}_2=\sum_{i=1}^{\mathcal{N}} \hat{n}_2 (x_i) $ the integrated signal from the spatial resolving detector. This alternative protocol is depicted by the dashed path in Fig. \ref{corr_scheme}. Note that  $\langle \hat{N}_1 \rangle/\langle \hat{N}_2 \rangle=\bar{t}$, where $\bar{t}$ is the average transmission of the object, i.e. $\bar{t}=(1/\mathcal{N})\sum_{i=1}^{\mathcal{N}} t (x_i)$. For DGI the reconstructed image is given by:

\begin{equation}
\begin{split}
S_{\mathrm{DGI}}(x_j)=\langle \delta \hat{N}_{\mathrm{\mathrm{DGI}}} \delta \hat{n}_2(x_{j})\rangle= S_{\mathrm{GI}}(x_j)- \bar{t} \langle \delta^2 \hat{n}_2(x_{j})\rangle
\end{split}
\end{equation}
In the last equality we have used the fact that modes collected by different pixels of the resolving detector are uncorrelated.
Making use of Eq. (\ref{var_n2}, \ref{cov_Th}, \ref{cov_SPDC}) we get
\begin{eqnarray}
S^{th}_{\mathrm{DGI}}(x_j)=n_2 \left[ \frac{n_2}{M} \delta t (x_j) -  \bar{t}      \right] \label{eqgdi}\\
S^{tw}_{\mathrm{DGI}}(x_j)=n_2 \left[\left( \frac{n_2}{M}+\eta\right) \delta t (x_j) -  \bar{t} (1-\eta)     \right]
\end{eqnarray}

From Eq. (\ref{eqgdi}), in the limit $n_2/M\gg1$, one retrieves the main feature of DGI as proposed in \cite{DGI}, namely
its sensitivity to the spatial change in the transmission of the object, $\delta t(x_j) = t(x_j) - \bar{t}$, rather than to the absolute value $t (x_j)$. This explains its significant advantages in the reconstruction of small or highly transparent objects. However, it is also clear that in the opposite regime of small number of photons detected per spatio temporal mode, the information on the transmission profile of the object is substantially lost. With twin beam, the transmission profile can be reconstructed even for $n_2/M\ll1$, provided that the efficiency $\eta$ is sufficiently high.
Anyway, the possibility to practically get a faithful image is determined by the SNR. It is important to reduce the noise as much as possible on the reconstruction. For this reason, following the method developed in \cite{moreau,absorption}, we propose a generalization of the DGI protocol by replacing  $\hat{N}_{\mathrm{\mathrm{DGI}}}$ with $\hat{N}_k = \hat{N}_1 - k \hat{N}_2$, leading to
\begin{equation}\label{signal_k}
S_{k}(x_j)=\langle \delta \hat{N}_k \delta \hat{n}_2(x_{j})\rangle= S_{\mathrm{GI}}(x_j)- k \langle \delta^2 \hat{n}_2(x_{j})\rangle
\end{equation}
where $k$ is a constant that can be set in order to minimize the noise in the estimation of $S_{k}(x_j)$, approximately given by:

\begin{eqnarray}\label{noise_k}
\delta^2 S_{k}(x_j)&\approx &\langle \delta^2 \hat{N}_k \rangle\langle\delta^2 \hat{n}_2(x_{j})\rangle \\\nonumber
\langle \delta^2 \hat{N}_k \rangle&= &\langle \delta^2 \hat{N}_1 \rangle+k^2\langle \delta^2 \hat{N}_2 \rangle-2k\langle \delta \hat{N}_1\delta  \hat{N}_2\rangle
\end{eqnarray}
For $k=0$, i.e. for conventional GI, the variance reduces to  $\delta^2 S_{\mathrm{GI}}(x_j)=\langle \delta^2 \hat{N}_1 \rangle\langle\delta^2 \hat{n}_2(x_{j})\rangle$ and making use of Eq. (\ref{var_n2}) and (\ref{var_n1}),
\begin{equation}\label{noise_GI}
\delta^2 S_{\mathrm{GI}}(x_j)=\mathcal{N} n_2^{2}\left(\bar{t}+\frac{n_2}{M} \bar{t^2}\right) \left(1 + \frac{n_2}{M}\right)
\end{equation}
 Eq. (\ref{noise_GI}) shows that in the limit of $n_2/M \gg 1$ we find the same noise dependence from the mean transmittance squared reported in \cite{DGI}, although the general expression predicts a linear dependence in the opposite regime.

For the sake of simplicity, we focus on the case of a two level object, with $\epsilon \mathcal{N}$ resolution cells of transmission $t_-$ and  $(1-\epsilon)\mathcal{N}$ of transmission $t_+$. Therefore, $\epsilon$ represents the fraction of the detection area occupied by the lower transmittance portion of the object. Note that $\bar{t}$ can be written in terms of  $0 \leq \epsilon \leq 1$ as $\bar{t}= (1-\epsilon) \mathcal{N} t_+ + \epsilon \mathcal{N} t_-$.
Under these assumptions the SNR is defined as:

\begin{equation}\label{SNR}
\begin{split}
\mathrm{SNR}=\frac{|\langle S_+\rangle - \langle S_- \rangle|}{\sqrt{ \delta ^2 \langle S_+\rangle + \delta^2 \langle S_- \rangle}},
\end{split}
\end{equation}
where  $\langle S_{\pm} \rangle $ and $ \delta ^2 \langle S_{\pm}  \rangle$ are respectively the mean value of the reconstructed image in correspondence of $t_{\pm}$ and its associated variance.
Considering $t_+=1$, $t_-=0$ and no electronic noise, the SNR in the thermal and  SPDC case for the GI protocol are:
\begin{equation}\label{SNRth}
\begin{split}
\mathrm{SNR}_{\mathrm{GI}}^{th}=\sqrt{H} \frac{1}{\sqrt{2 \mathcal{N}(1-\epsilon)}} \frac{n_2}{n_2+M},
\end{split}
\end{equation}
\begin{equation}\label{SNRpdc}
\begin{split}
\mathrm{SNR}_{\mathrm{GI}}^{tw}=\sqrt{H} \frac{1}{\sqrt{2 \mathcal{N}(1-\epsilon)}} \frac{n_2+M \eta}{n_2+M}
\end{split}
\end{equation}
Note that in this case $\epsilon$ simply becomes the fraction of the detection area occupied by the object. In the SNR expression the factor $\sqrt{H}$, being \emph{H} the number of frames used to estimate $S$, simply comes from the central limit theorem, where the uncertainty on a mean value scales as $\sqrt{H}$, e.g. $\delta ^2 \langle S(x_j) \rangle = \delta ^2  S(x_j) / H $. From Eq.s (\ref{SNRth})-(\ref{SNRpdc}) it emerges one of the weakness of the conventional GI, namely that for small $\epsilon$ the SNR drops down, making difficult the reconstruction of small objects in the field of view.

\begin{figure}[htbp]
	\centering
	\includegraphics[ trim= 4cm 3cm 3cm 0cm, clip=true, width=0.45\textwidth, angle=0]{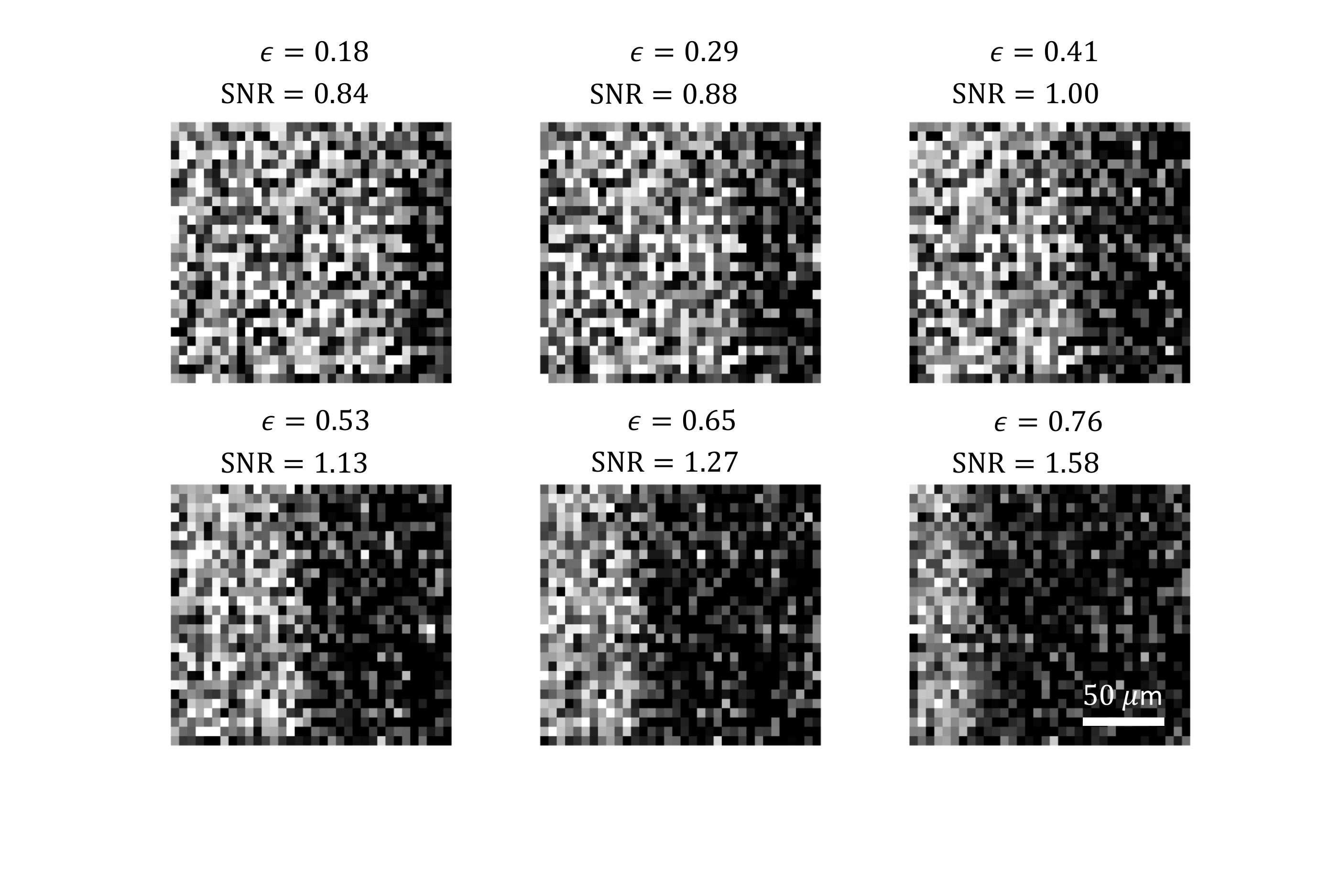}
	\caption{Experimental GI reconstructions of a binary object ($t_+=1$, $t_-=0$) for  $\mathcal{N}=952$ and $H=2000$, using SPDC light. The object consists in a totally absorbing deposition occupying a variable fraction $\epsilon$ of the reconstructed area. Different values of $\epsilon$ are considered and the corresponding SNR estimated; for small $\epsilon$, e.g. $\epsilon=0.18$, SNR$<$1 and the deposition on the right side is almost hidden in the noise.}\label{confronto_ricostruzioni_2000_frames}
\end{figure}

In Fig. \ref{confronto_ricostruzioni_2000_frames} we report an example of GI reconstruction for a two level object, for different values of $\epsilon$. From these reconstructions it is possible to experimentally estimate the SNR. $\langle S_{\pm} \rangle$ and $ \delta^2 \langle S_{\pm} \rangle$ are estimated as the mean value and the variance on the reconstructed images, in correspondence of the regions of transmittance $t_{\pm}$ respectively. It can be appreciated how the object, a totally absorbing deposition on the right side of the field of view, better emerges from the noise as $\epsilon$ increases.

This issue of the conventional GI can be overcome by considering the generalized protocol $S_k(x_j)$. Indeed, from Eq. (\ref{noise_k}) we argue that exploiting the spatial correlations between the two beams, i.e. $\langle \delta \hat{N}_1 \delta \hat{N}_2 \rangle \neq 0$, and choosing $k$ appropriately, the noise can be reduced. This idea is at the basis of the optimization procedure we implemented. In particular, we maximize the SNR respect to $k$, and define the ODGI protocol as:
\begin{equation}\label{SOtt}
\begin{split} S_{\mathrm{ODGI}}(x_j)=S_{k_{opt}}(x_j)=\langle \delta \hat{N}_{opt} \delta \hat{n}_2(x_{j})\rangle ,\\ \hat{N}_{opt} = \hat{N}_1 - k_{opt} \hat{N}_2.
\end{split}
\end{equation}
%In general $k_{opt}$ is function of the mean object transmittance $\bar{t}$, of the source brightness as well as $n_2$, $\eta$ and $\Delta_{el}$.
The general expression for $k_{opt}$, in the twin-beam case, is:
\begin{equation}
    k_{opt}^{tw}=\frac{n_2 (n_2 + M \eta) \bar{t}}{n_2^2+M (n_2+\Delta_{el}^2)}
\end{equation}
while for the thermal case a similar expression holds but without the term $M \eta$ at the numerator.
%It can be simplified considering the two opposite regimes of interest:
Focusing on the twin-bem case, we have:
\begin{eqnarray}
	n_2/M \gg 1&:& k_{opt}^{tw}=\bar{t}= \frac{\langle \hat{N}_1 \rangle}{\langle \hat{N}_2 \rangle} \label{eq20}\\
	n_2/M \ll \eta&:& k_{opt}^{tw}= \frac{n_2}{n_2+\Delta_{el}^2} \eta \cdot \frac{\langle \hat{N}_1 \rangle}{\langle \hat{N}_2 \rangle} \label{eq21}
\end{eqnarray}
Eq. (\ref{eq20}) shows that for high number of detected photons per mode %, typical of experiments with thermal light,
$S_{\mathrm{ODGI}}$ coincides with $S_{\mathrm{\mathrm{DGI}}}$. Remarkably, in this case $k_{opt}$  does not depend on experimental imperfections.

For low brightness sources, as shown in Eq. (\ref{eq21}), $S_{\mathrm{\mathrm{DGI}}}$ coincides with $S^{tw}_{\mathrm{ODGI}}$ only in the ideal case of $\eta =1$ and $\Delta_{el}=0$; in all other cases ODGI performs better. However, in order to evaluate $k^{tw}_{opt}$, absolute values of the channel efficiency $\eta$ and the detection noise $\Delta_{el}$ should be known \cite{absorption,alessio}.

\begin{figure}
	\centering
	\subfigure[]{
		\label{fig:first}
		\includegraphics[trim= 2cm 0cm 0cm 0cm, clip=true, width=0.47\textwidth]{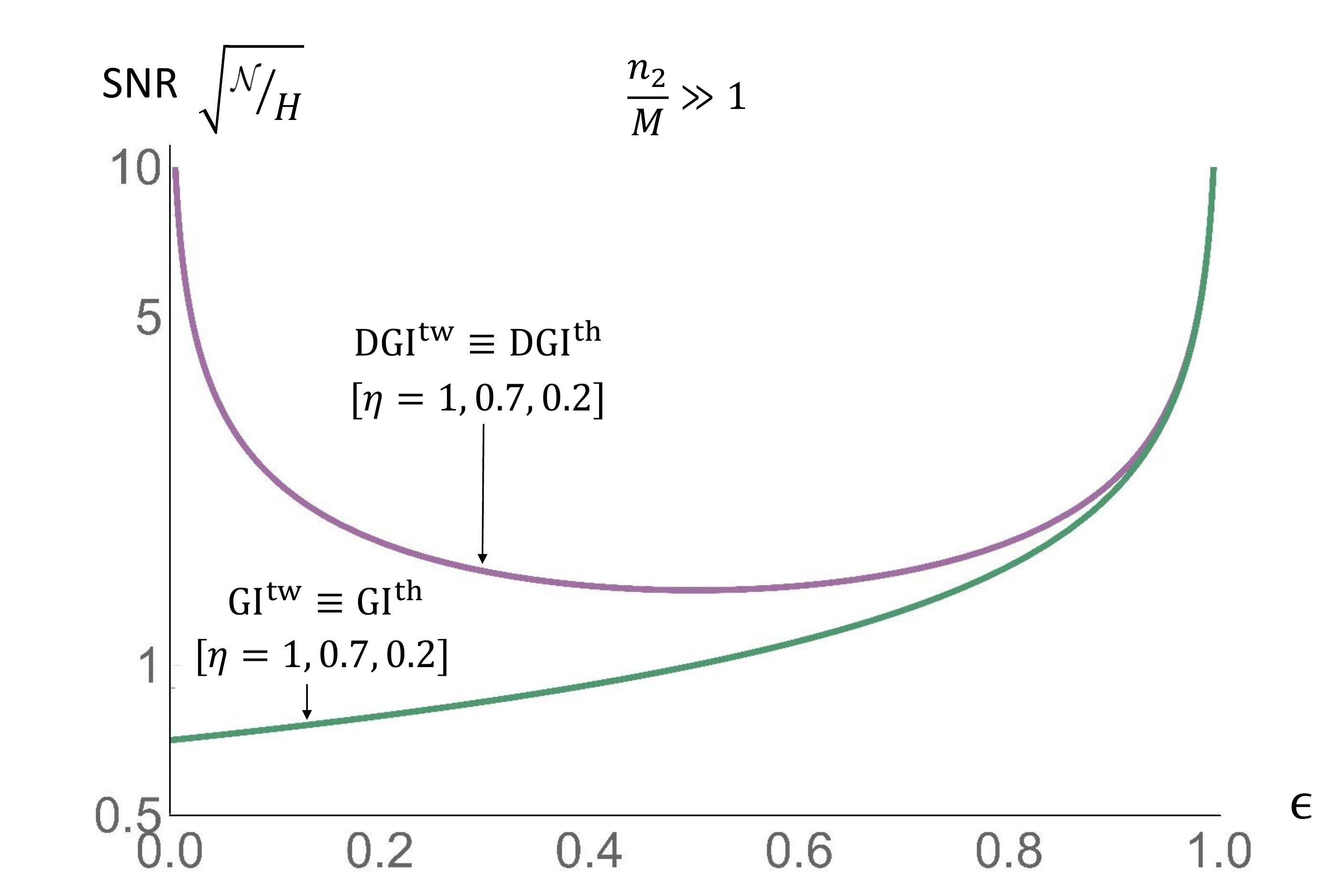}}
	\qquad
	\subfigure[]{
		\includegraphics[trim= 2cm 0cm 0cm 0cm, clip=true, width=0.47\textwidth]{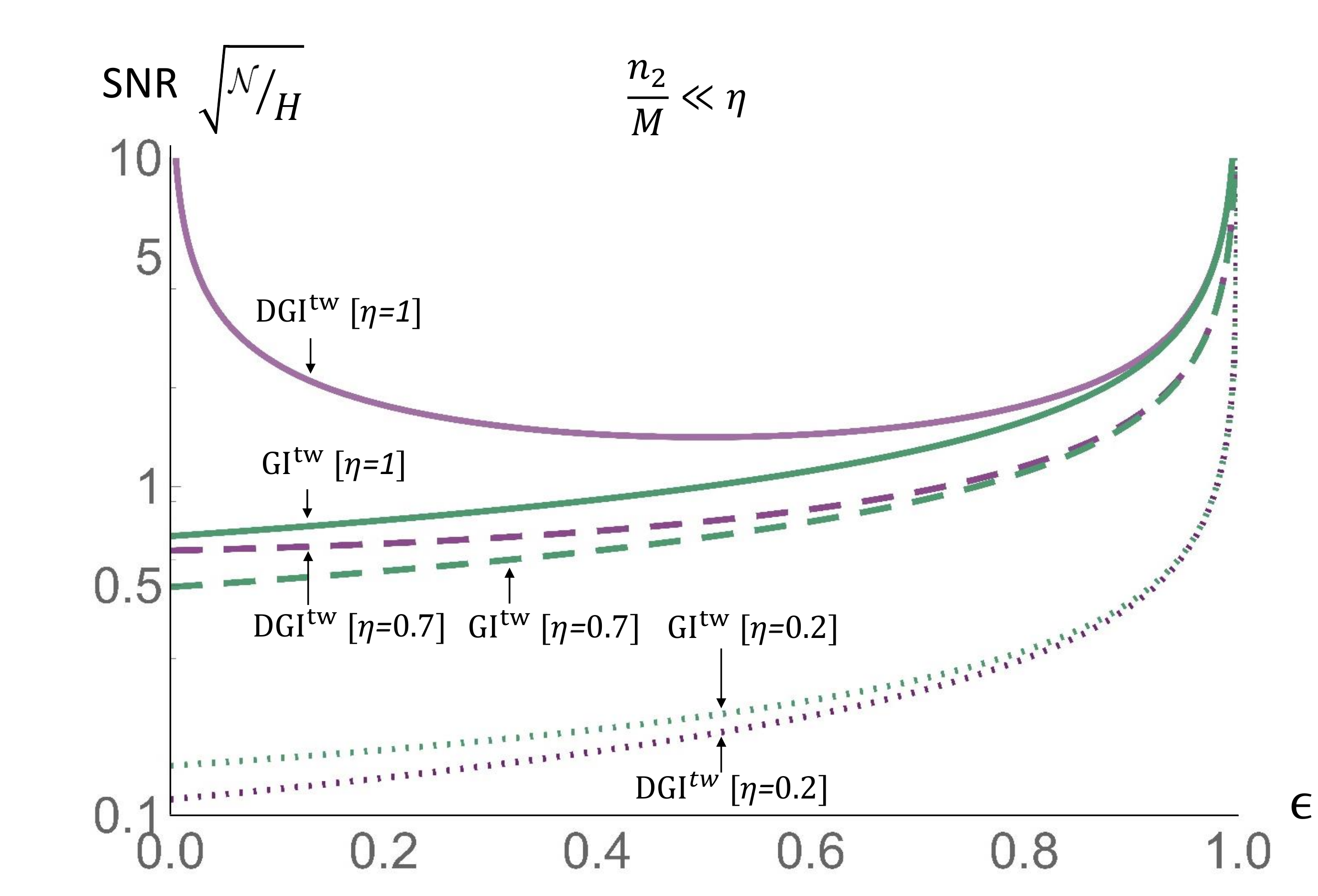}}
	\caption{SNR, normalized per the number of frames $H$ and the number of pixels in the reconstructed area $\mathcal{N}$, in function of the object occupation fraction $\epsilon$, for different values of the channel efficiency $\eta$ ($t_+=1$, $t_-=0$, $\Delta_{el}=0$).
	(a) High brightness regime. (b) Low brightness regime, twin-beam case.} \label{T2role}
\end{figure}

The improvement offered by DGI versus the conventional GI protocol can be quantified in terms of SNR. For the sake of simplicity we report the expressions neglecting the  electronic noise, namely $\Delta_{el}^2 \ll n_2$, and considering a binary object ($t_+=1$, $t_-=0$):

\begin{eqnarray}
	 n_2/M \gg 1/\epsilon &:& \frac{\mathrm{SNR}^{tw}_{\mathrm{DGI}}}{\mathrm{SNR}^{tw}_{\mathrm{GI}}}=\frac{\mathrm{SNR}^{th}_{\mathrm{DGI}}}{\mathrm{SNR}^{th}_{\mathrm{GI}}}=\frac{1}{\sqrt{\epsilon}} \label{ratio2} \\
	 n_2/M \ll 1 &:& \frac{\mathrm{SNR}^{tw}_{\mathrm{DGI}}}{\mathrm{SNR}^{tw}_{\mathrm{GI}}}=\frac{1}{\sqrt{2(\eta-\frac{1}{2})(\epsilon-1)+1}}\label{SNR_ratio}
\end{eqnarray}
These results can be graphically appreciated in Fig. \ref{T2role}. In the high-intensity regime, corresponding to Eq. (\ref{ratio2}) and reported in Fig. \ref{T2role}a, SPDC and thermal light show the same performance and the SNRs for both GI and DGI are independent from the channel efficiency. Note that the validity condition of Eq. (\ref{ratio2}) requires higher and higher values of detected photons per mode as $\epsilon$ decreases.
For small $\epsilon$ the DGI advantage is relevant, an extremely interesting feature in view of real applications. %Moreover, as can be appreciated in Fig. \ref{T2role}(a), not only the advantage of DGI over GI is unaffected by losses, but also the SNR itself, according to Eq. (\ref{SNRth}).

On the contrary, losses in the low intensity regime have a significant impact. From Eq. (\ref{SNR_ratio}) comes out that, despite we find the same result of Eq. (\ref{ratio2}) for $\eta=1$,  the improvements offered by DGI over GI decreases with losses. In particular, for $\eta<1/2$, DGI performs even worse than the conventional protocol. The dependence of the SNR from the channel efficiency, at low brightness, is evident in Fig. \ref{T2role}(b).

The ODGI protocol can partially compensate for losses, in particular:

\begin{equation}
    n_2/M \ll \eta: \frac{\mathrm{SNR}^{tw}_{\mathrm{ODGI}}}{\mathrm{SNR}^{tw}_{\mathrm{GI}}}=\frac{1}{\sqrt{\eta^2(\epsilon-1)+1}}\label{snr_odgi}
\end{equation}
From Eq. (\ref{snr_odgi}) it emerges that, for low brightness sources, ODGI performs always better than the conventional GI, but it is also permanently better than DGI, as it emerges  comparing Eq. (\ref{snr_odgi}) with Eq. (\ref{SNR_ratio}). Note that in these equations we don't report the expressions for thermal light since in this regime thermal correlations are order of magnitude weaker than for twin-beam light (see Eq. (\ref{eq7})), thus requiring an unfeasable number of frames for achieving a sufficiently high SNR.

\begin{figure}[htbp]
	\centering
	\includegraphics[ trim= 1cm 1cm 0cm 1cm, clip=true, width=0.47\textwidth]{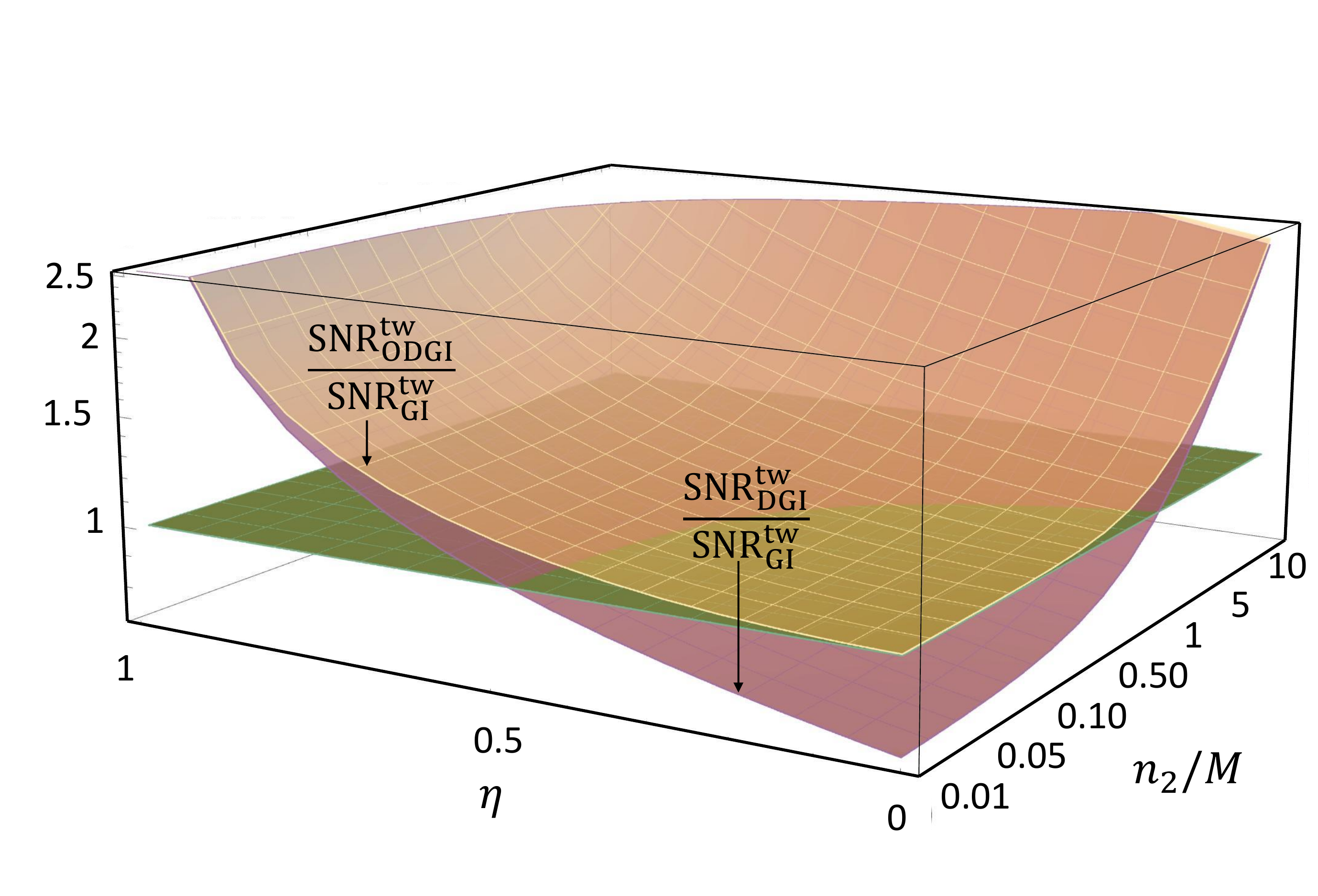}
	\caption{DGI and ODGI are compared in terms of SNR with GI, at varying both the channel efficiency $\eta$ and the number of detected photons per mode $n_2/M$. The surfaces are obtained for the twin-beam case, considering a binary object ($t_{+}=1, t_{-}=0$) and $\epsilon=0.1$.}\label{ratio3d}
\end{figure}

The general performance of DGI and  ODGI with respect to conventional GI, for the SPDC case, is reported in Fig. \ref{ratio3d} in function of the channel efficiency $\eta$ and the number of detected photons per mode $n_2/M$, fixing $\epsilon=0.1$. Considering the front plane, corresponding to  $n_2/M=0.01$, one can observe that the highest ODGI advantage over both the other protocols occurs when $\mathrm{SNR}^{tw}_{\mathrm{DGI}}=\mathrm{SNR}^{tw}_{\mathrm{GI}}$, i.e. for  $\eta \sim 0.5$.
Moreover, it emerges that increasing the brightness of the source allows obtaining an advantage with respect to GI for lower value of the channels efficiency $\eta$, so that, depending on the experimental condition, the optimal trade-off can be designed.

\section*{Experiment}

Here we present an experiment comparing GI, DGI and ODGI protocols in the low brightness regime using twin-beam generated by SPDC.

\begin{figure}[htbp]
	\centering
	\includegraphics[ trim= 1.5cm 4cm 8cm 7cm, clip=true, width=0.49\textwidth, angle=0 ]{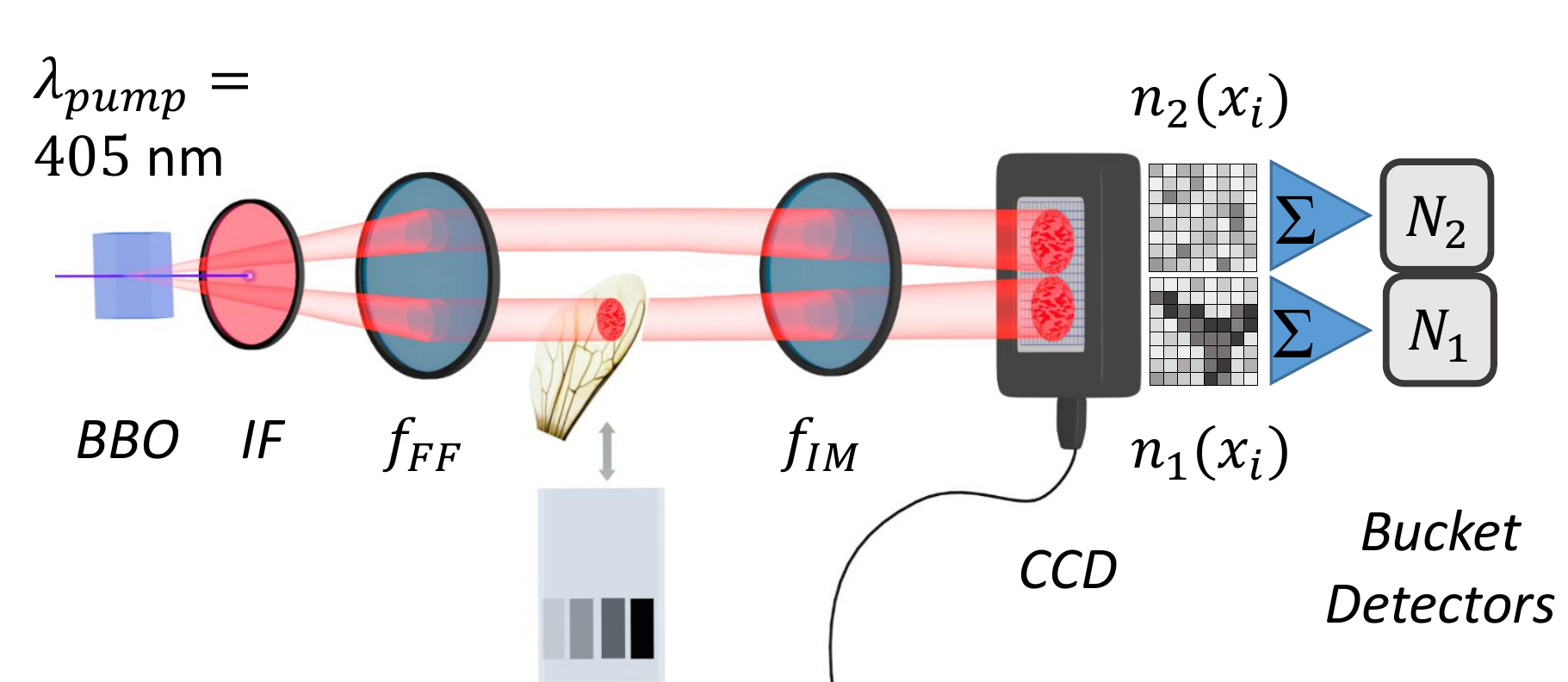}
	\caption{Scheme of the experimental set-up. In the BBO crystal two beams with perfect correlation in the photon number (twin-beam state) are generated.  The probe beam interacts with the sample and is detected by a half of a CCD camera chip, while the other beam goes directly to the second half of the chip. The equivalent bucket detector on both channels are simulated by integrating (summing pixels signal) over the two regions. Two different samples are used: a set of uniform depositions of different transmittance $t_-$ and a biological object, i.e. a wasp wing.}\label{setup}
\end{figure}

The experimental set-up is reported in Fig. \ref{setup}.
A twin-beam state with degeneracy wavelength $\lambda_1=\lambda_2= 810 \mathrm{nm}$ is produced pumping a 1cm Type-II-Beta-Barium-Borate (BBO) non linear crystal with a CW laser-beam (100mW  at $\lambda_{p}=405 \mathrm{nm})$. In this process, with a certain small probability, a photon of the pump is down-converted into two photons.  The momentum conservation implies that the two downconverted photons emerge with opposite transverse momenta $\textbf{q}_{1}=-\textbf{q}_{2}$. In the far field of the emission, realized at the focal plane of a lens with $f_{FF}=1$cm, momentum correlation is mapped in position correlation, $\textbf{x}_1=-\textbf{x}_2$, with  $\textbf{x}_i=\frac{\lambda_i \textbf{q}_i f_{FF} }{2 \pi}$.
To detect only photons around the degeneracy and to cut the pump, an interference filter with 20nm bandwidth centered at 800nm  is placed after the crystal.
To take advantage of the point-to-point far-field correlations, the object is placed directly  at the focal plane of the $f_{FF}$  lens , and a second lens with focal length $f_{IM}=1.6$cm, is used to image this plane to the camera sensor, with a magnification factor $M=7.8$.

The detector is a charge-coupled-device (CCD) camera Princeton Inst. Pixis 400BR Excelon operating in linear mode, i.e. the signal is proportional to the incident number of photon (intensity) in the acquisition time, and cooled down to $-70^\circ$C. It presents high quantum efficiency, nominally $>95\%$ at 810nm, 100$\%$ fill factor and low noise (read-noise has been estimated to be $\Delta_{el}=5 e^-/(pixel \cdot frame)$ at 100kHz acquisition rate, and $\Delta_{el}=13 e^-/(pixel \cdot frame)$ when the higher 2MHz digitization rate is used). The detector area measures $(13.3 \mathrm{mm})^2$ and the size of a physical pixel is 13 $\mu \mathrm{m}$, nevertheless for this experiment we use $3 \times 3$ hardware binning.
It is known that the ultimate resolution achievable in ghost imaging is given by the second order coherence area of the light at the object plane. In particular, following the procedure in \cite{ccdcalibration}, we estimated this coherence area to be $(5\mu \mathrm{m})^2$. The binned pixel size corresponds roughly to the coherence area, taking into account the 7.8 magnification factor from the object plane to the detection plane, so that pixels collect independent optical mode thus remaining statistically uncorrelated to each other.
Note that in our experimental configuration there is no physical bucket detector. The equivalent bucket detectors on both the beams are simulated by integrating over the respective regions of the sensor.

The average number of photons detected per pixel per frame is $ n_2 \sim 10^3$ and the corresponding number of modes is $M = M_{sp} M_{temp}$, where $M_{sp}$ and $M_{temp}$ are the number of spatial and temporal modes.  Being the acquisition time of a single frame 50 ms, and the coherence time of the SPDC process around $10^{-12}$s, it follows $M_{temp} \sim 5 \cdot 10^{10}$. The dimension of a binned pixel is chosen in order to have $M_{sp} \sim 1$ incident. It follows that the condition of very low brightness is fulfilled: $n_2/ M \sim 2 \cdot 10^{-8} \ll \eta$.

In order to demonstrate the non-classicality of the detected intensity correlation, we evaluate the NRF parameter defined as $\mathrm{NRF}=\langle \Delta^{2}(\hat{n}_1- \hat{n}_2)\rangle/\langle \hat{n}_1+\hat{n}_2 \rangle$, that is the variance of the photon number difference between a pair of correlated pixels normalized by their sum (the shot noise bound). Only non-classical correlation allows to have $0<\mathrm{NRF}<1$ \cite{review2017, Ruo19}. In our case we estimated $\mathrm{NRF}=0.77 \pm 0.02$. Note that NRF, in our setup, depends on the pixel size with respect to the coherence area, thus it is possible to reach lower value of NRF integrating the signals over larger areas. However, this is not the scope of the present work.

To experimentally validate the theoretical model we image four different objects, each of them presenting two levels of transmittance ($t_{+} \sim 1$ for all cases, $t_{-}=0,0.25,0.34,0.52$ respectively). These samples consist in a AR-coated glass-slide with thin metallic depositions. For each sample $3\cdot 10^4$ frames are acquired at 100 kHz acquisition rate. In the data processing, a cropped region of $\mathcal{N}=28 \times 34$ binned pixels is reconstructed, including each time an increasing fraction $\epsilon$ of the low transmittance part of the object. The reconstruction is performed using the three different protocols $S_{\mathrm{GI}}$, $S_{\mathrm{DGI}}$, $S_{\mathrm{ODGI}}$. An example of the corresponding reconstructions is presented in Fig. \ref{confronto_ricostruzioni_2000_frames} in case of the conventional GI. The dependence of SNR, in Eq. (\ref{SNR}),  from  $\epsilon$ is evaluate performing spatial statistics on this kind of images.

\begin{figure}[htbp]
	\centering
	\includegraphics[ trim= 8cm 0cm 6.5cm 0cm, clip=true, width=0.47\textwidth ]{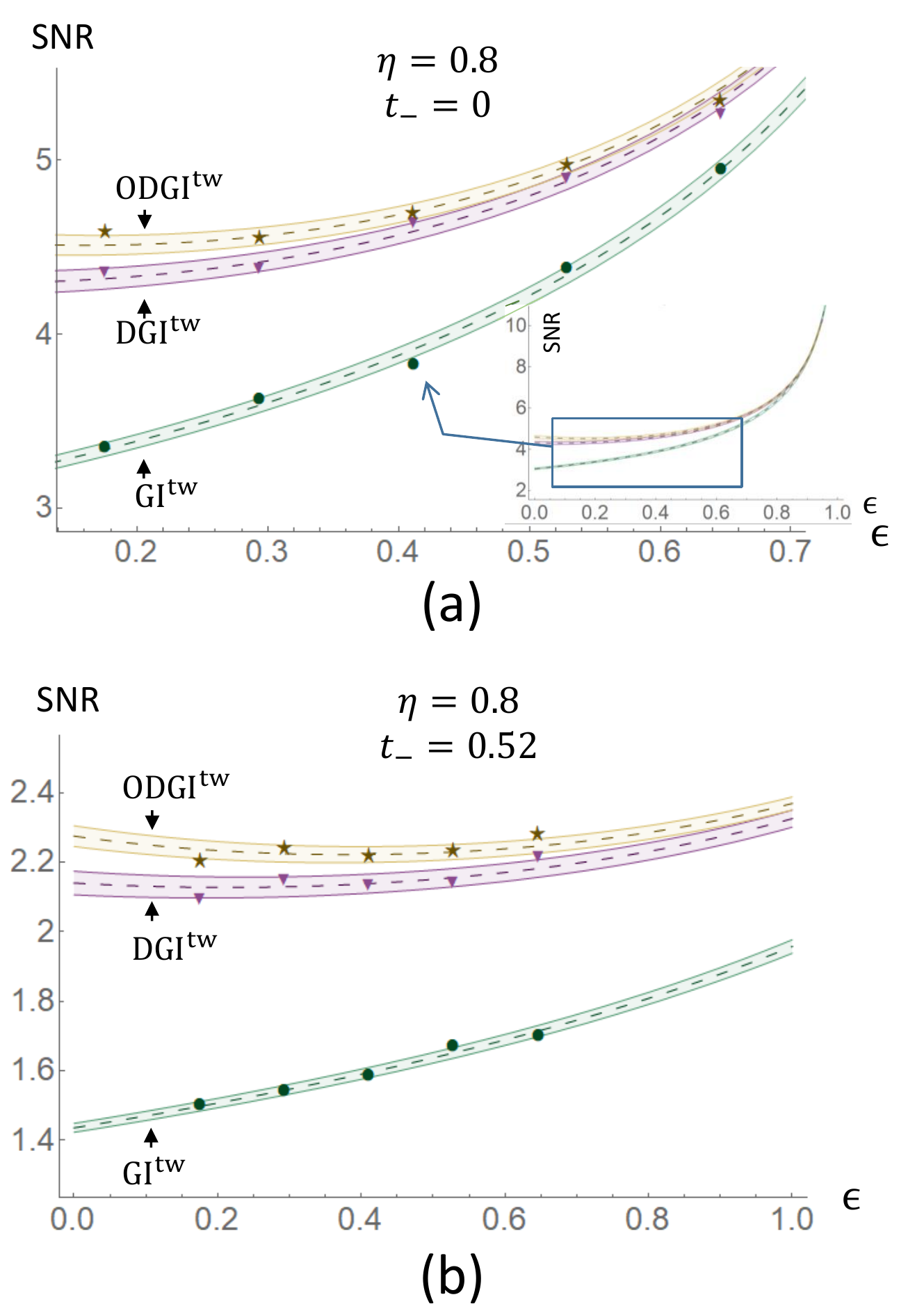}
	\caption{SNR in function of $\epsilon$, the fraction of the detection area occupied by the deposition, for (a) $t_-=0$ and (b) $t_-=0.52$. Green color refers to GI, purple to DGI and yellow to ODGI. The dots are the experimental data, obtained for  $H \sim 3 \cdot 10^4$ frames and an ares of $\mathcal{N}=952$ pixels. The dashed lines are obtained fitting the data with the theoretical model, considering $\eta$ as free parameter. The confidence region at $1\sigma$ is also reported as colored bands around the curves.}\label{ass}
\end{figure}

Fig. \ref{ass} (a) and (b) report the results for two different transmittances, $t_{-}=0$ and $t_{-}=0.52$ respectively. The dashed lines are obtained fitting the experimental data with the theoretical model, considering $\eta$ as free parameter.

%\begin{figure}
%	\centering
%	\subfigure[]{
%		\label{ass100}
%		\includegraphics[trim= 0cm 0cm 0cm 0.9cm, clip=true,width=0.47\textwidth]{test100.pdf}}
%	\qquad
%		\subfigure[]{
%		\label{ass050}
%		\includegraphics[ trim=0cm 0cm 0cm 0.9cm, clip=true,width=0.47\textwidth]{test050.pdf}}
%	\caption{SNR in function of $\epsilon$, the fraction of the detection area occupied by the deposition, for (a) $t_-=0$ and (b) $t_-=0.52$. Green color refers to GI, purple to DGI and yellow to ODGI. The dots are the experimental data, obtained for  $H \sim 3 \cdot 10^4$ frames and an ares of $\mathcal{N}=952$ pixels. The dashed lines are obtained fitting the data with the theoretical model, considering $\eta$ as free parameter. The confidence region at $1\sigma$ is also reported as colored bands around the curves.}\label{ass}
%\end{figure}

\begin{figure}[htbp]
	\centering
	\includegraphics[ trim= 1cm 0cm 0cm 0cm, clip=true, width=0.47\textwidth ]{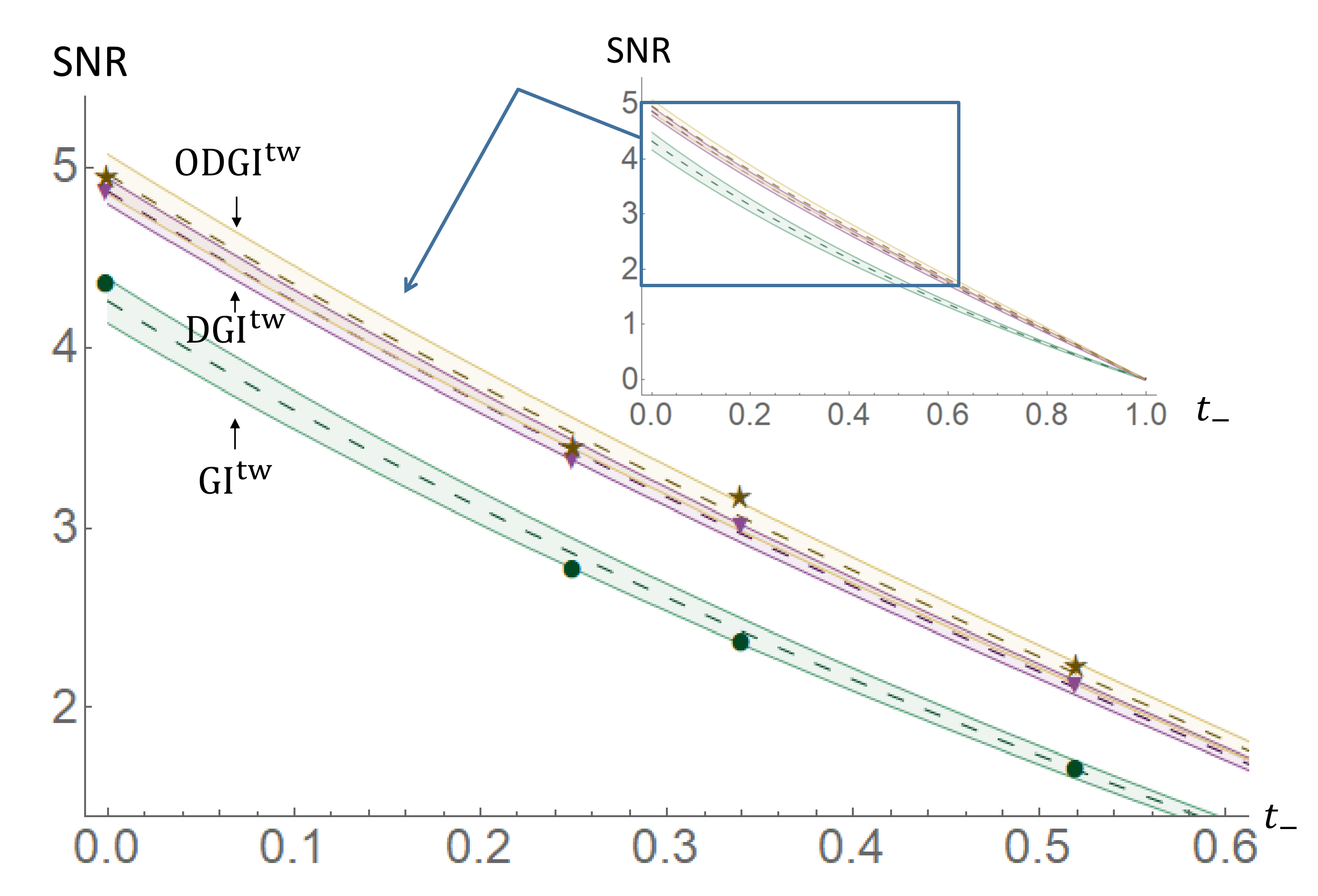}
	\caption{SNR in function of $t_-$, fixed $\epsilon=0.52$. Green color refers to GI, purple to DGI and yellow to ODGI. The dots are the experimental data, obtained for  $H \sim 3 \cdot 10^4$ frames and an area of $\mathcal{N}=952$ pixels. The dashed lines are obtained fitting the data with the theoretical model, considering $\eta$ as free parameter. The confidence region at $1\sigma$ is also reported as colored bands around the curves.}\label{SNRvsT1m}
\end{figure}

In Fig. \ref{SNRvsT1m}, at fixed $\epsilon=0.52$, the SNR for different $t_{-}$ is reported. Also, in this case data are fitted using the theoretical curves with the efficiency $\eta$ as free parameter.

\begin{figure}[htbp]
	\centering
	\includegraphics[ trim= 0cm 0cm 0cm 0cm, clip=true, width=0.47\textwidth ]{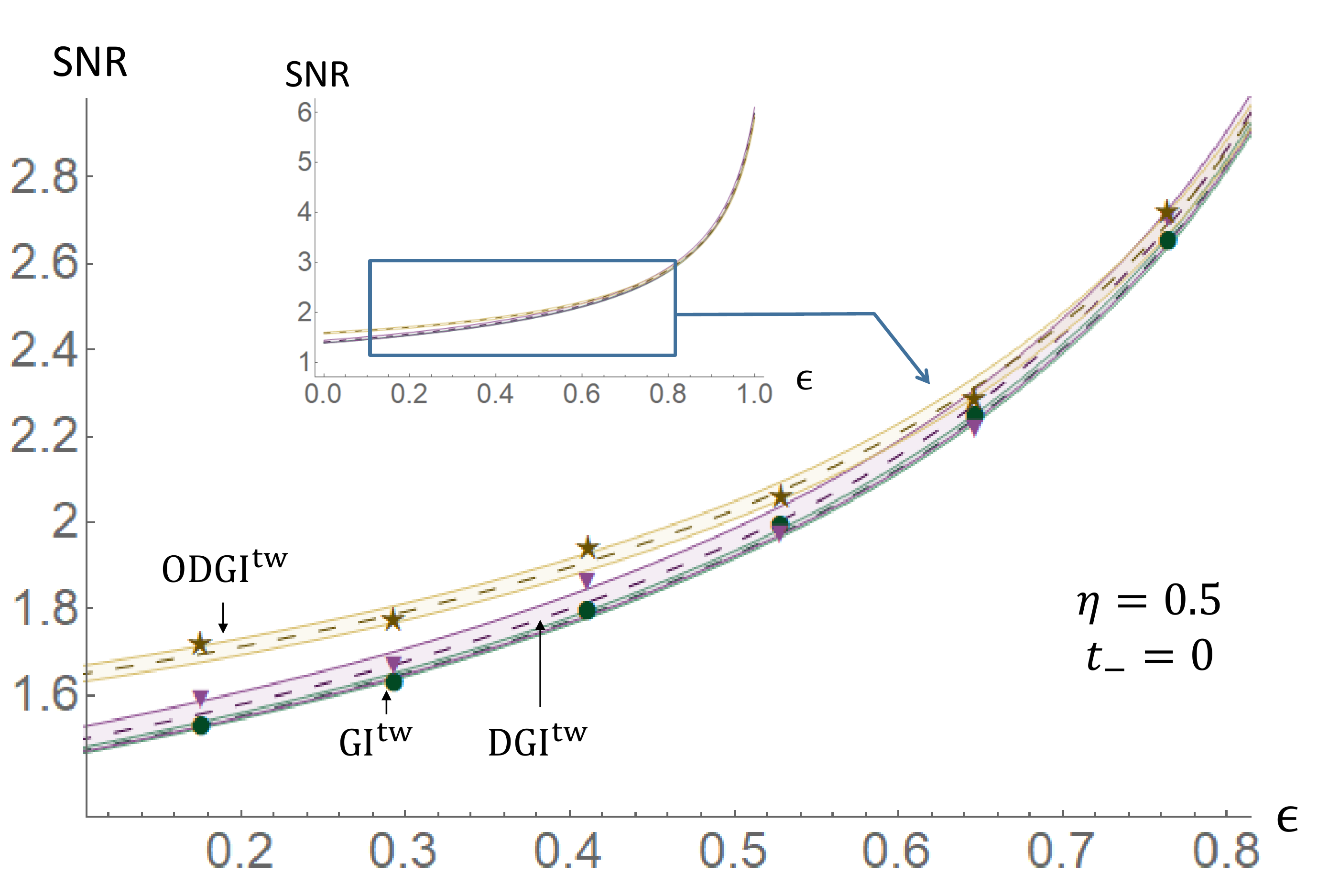}
	\caption{SNR in function of $\epsilon$, for $\eta \sim 0.5$. Green color refers to GI, purple to DGI and yellow to ODGI. The dots are the experimental data, obtained acquiring $ H \sim 3 \cdot 10^4$ frames and considering an area of $\mathcal{N}=952$ pixels. The dashed lines are obtained fitting the data with the theoretical model, considering $\eta$ as free parameter. The confidence region at $1\sigma$ is also reported as colored bands around the curves.}\label{eta05}
\end{figure}

\begin{figure}[htbp]
	\centering
	\includegraphics[ trim= 0cm 0cm 0cm 0cm, clip=true, width=0.47\textwidth ]{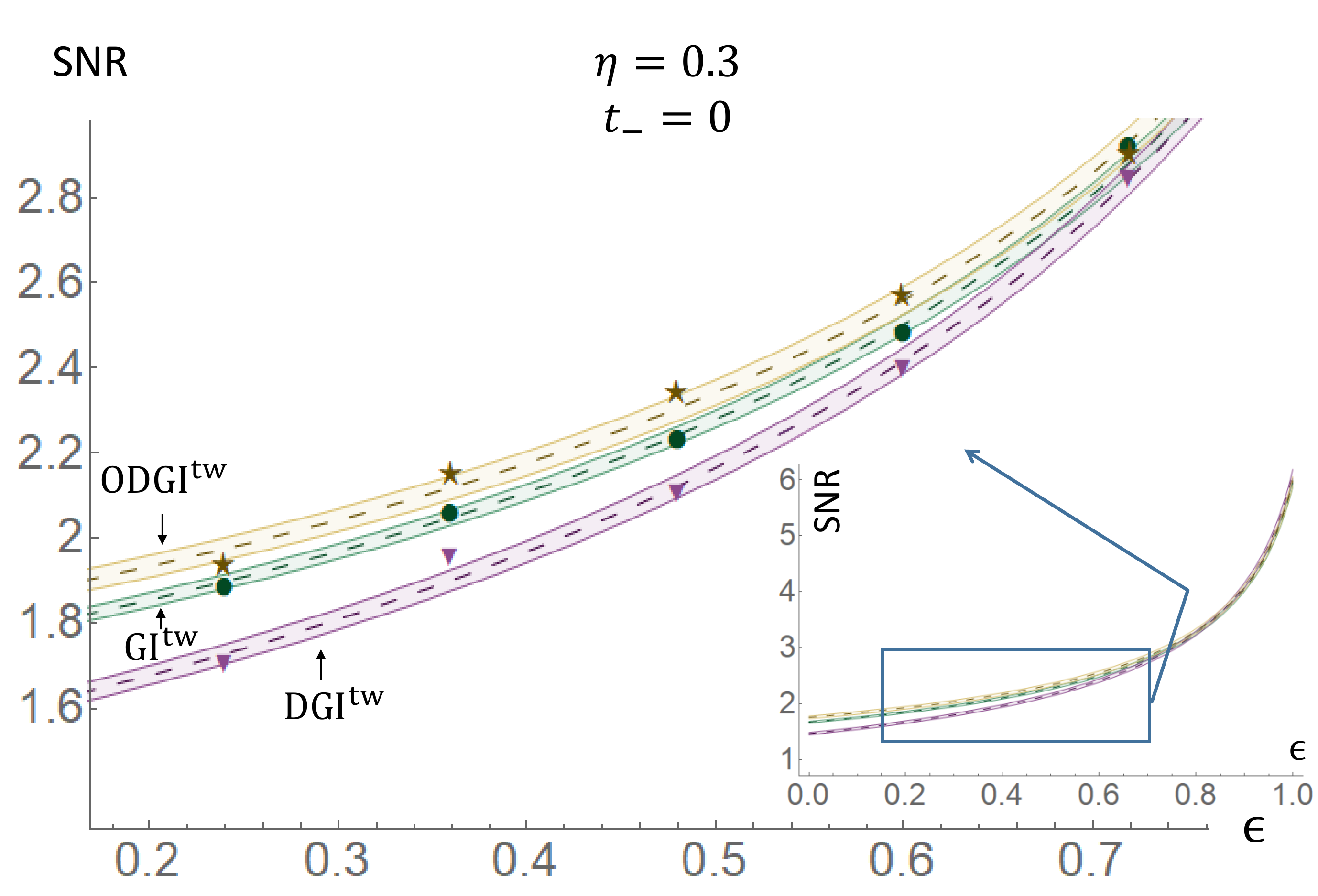}
	\caption{SNR in function of $\epsilon$, for $\eta \sim 0.3$. Green color refers to GI, purple to DGI and yellow to ODGI. The dots are the experimental data, obtained acquiring $ H \sim 3.5 \cdot 10^4$ frames and considering an area of $\mathcal{N}=525$ pixels. The dashed lines are obtained fitting the data with the theoretical model, considering $\eta$ as free parameter. The confidence region at $1\sigma$ is also reported as colored bands around the curves.}\label{eta03}
\end{figure}

For comparing the three protocols in presence of higher losses a neutral filter is placed on the twin-beam path. Two different values of channel efficiency are considered: $\eta \sim 0.5$ and $\eta \sim 0.3$. The results corresponding to these experimental situations are reported in Fig.s \ref{eta05} and \ref{eta03} respectively. In particular for $\eta = 0.3$, the DGI protocol performs worse than the conventional one, while the ODGI always offers an advantage.

In all the cases considered, see Fig.s \ref{ass}-\ref{SNRvsT1m}-\ref{eta05}-\ref{eta03}, the curves fit the experimental data properly, falling almost all the data in the $1\sigma$ confidence region.
A further element of consistency of the model is the accordance between the value of $\eta$ obtained from the fit and the one independently estimated with the absolute technique described in \cite{alessio, ccdcalibration, brida06, etaeval}, which extends the Klyshko method \cite{klyshko2, klyshko3}. Indeed, referring for example to Fig. \ref{ass} (a), the values of $\eta$ obtained from the fit and their standard uncertainty are:
	$\eta_{\mathrm{GI}}=0.798\pm 0.004$, $\eta_{\mathrm{DGI}}=0.786\pm 0.003$, $\eta_{\mathrm{ODGI}}=0.786\pm 0.003$, being the value independently estimated $\eta=0.794\pm 0.003$. The values are compatible with a confidence level of 95\%. Analogous results are obtained in all the other fits.

\begin{figure}[htbp]
	\centering
	\includegraphics[ trim= 0cm 3cm 2cm 4.5cm, clip=true, width=0.49\textwidth]{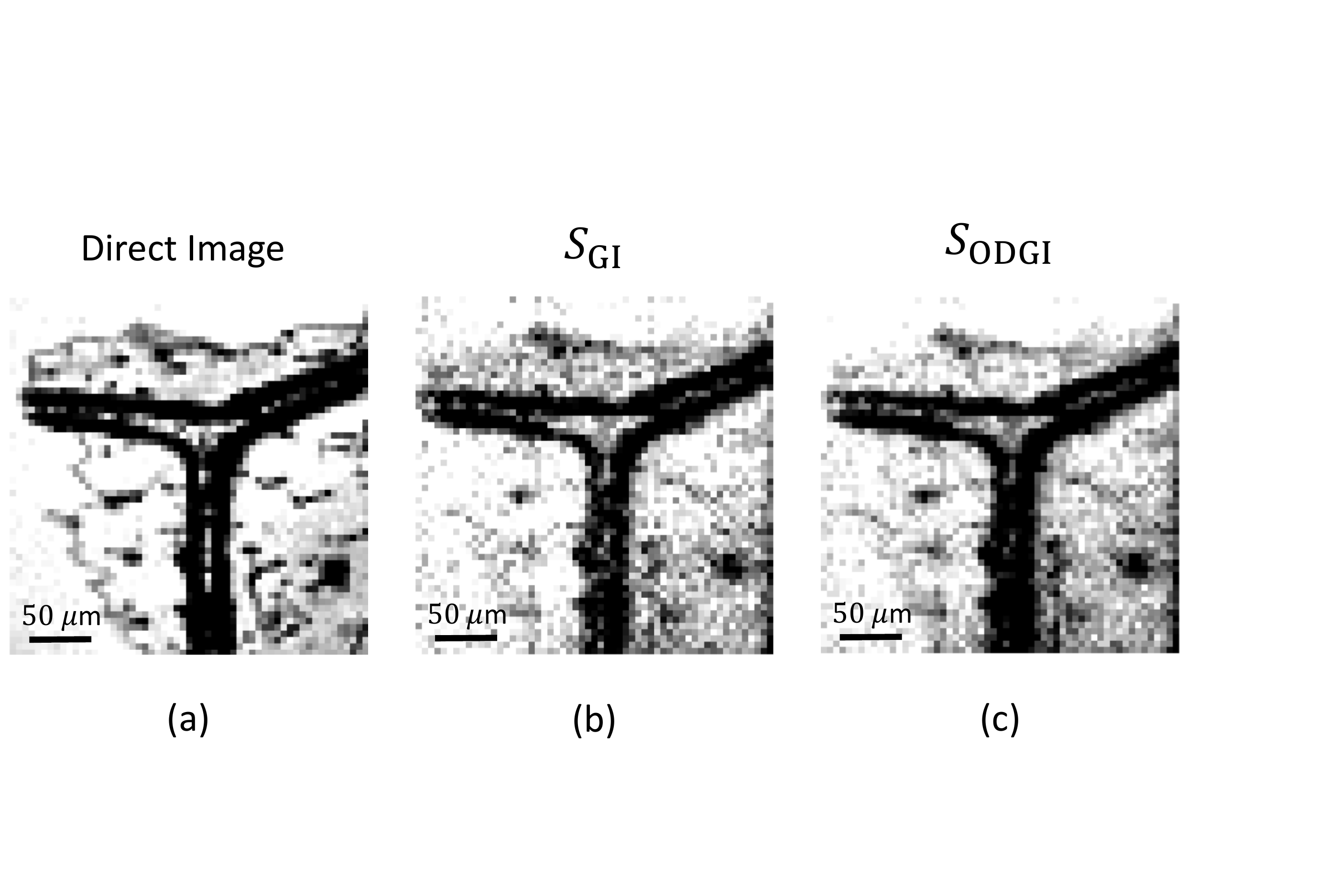}
	\caption{Images of a wasp wing with spatial resolution of $(5\mu \mathrm{m})^2$. (a) direct image, obtained averaging 5000 frames. (b)-(c) reconstruction using the GI and ODGI protocol respectively. The total region is divided into 9 sub-region, the protocol is applied to each of them and finally the complete image is recovered. 40 block of $5000$ images acquired at 2MHz are processed.}\label{wasp_wing}
\end{figure}

Finally, in order to demonstrate that our system can be interesting in view of real application, two different biological samples are imaged, in particular, a \emph{polistes} wasp wing and a green bug wing. The details cover a region of $\mathcal{N}=57\times 57$ binned pixels, corresponding to an area of $(285 \mu \mathrm{m})^2$ in the object plane. The resolution achieved with our set-up is $5 \mu$m, corresponding to the size of the second order coherence area at the object plane reached in the actual setup. The resolution is not the main concerns of our paper, focused on the SNR, but we mention that it is aligned with other recent ghost microscopy experiment \cite{Aspden15}. In Fig. \ref{wasp_wing} the \emph{polistes} wasp wing reconstructions, obtained for $2 \cdot 10^5$ frames acquired at 2MHz, are reported.
\begin{figure}[htbp]
	\centering
	\includegraphics[trim= 2cm 0cm 8cm 0cm, clip=true, width=0.49\textwidth]{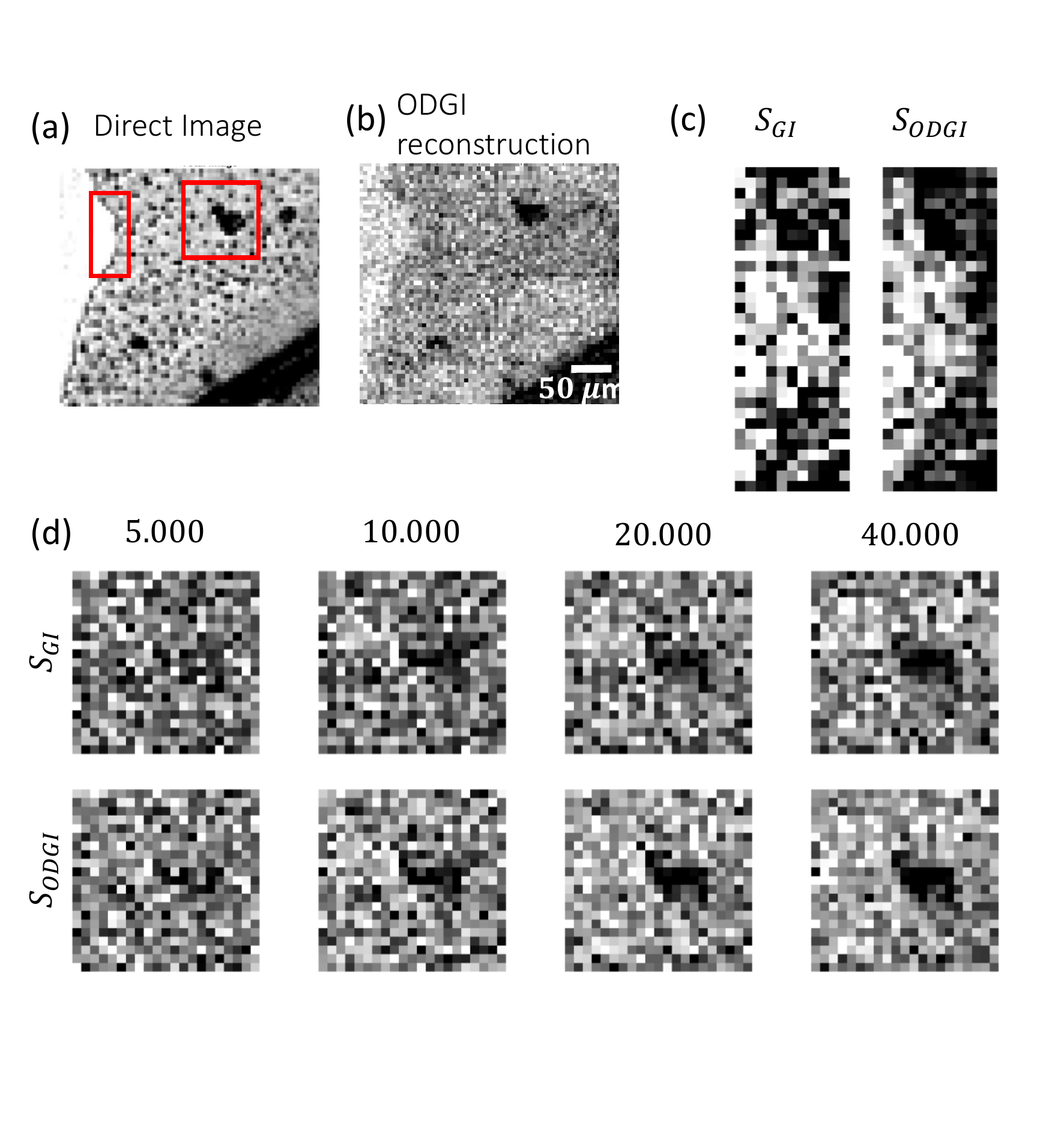}
	\caption{Images of a green bug wing with spatial resolution of $(5\mu \mathrm{m})^2$. (a) direct image, obtained averaging 5000 frames. (b) reconstruction using the ODGI protocol. The total region is divided into 9 sub-region, the protocol is applied to each of them and finally the complete image is recovered. 8 blocks of $5000$ images acquired at 100 kHz are processed. (c) $S_{GI}$ and $S_{ODGI}$ reconstruction of the border detail squared in (a). (d) $S_{GI}$ and $S_{ODGI}$ reconstruction of the spot detail squared in (a). The reconstruction for different number of frames is reported.}\label{detail}
\end{figure}
Fig. \ref{wasp_wing}(a) is the direct image. To obtain the reconstructed images via $S_{\mathrm{GI}}$ and $S_{\mathrm{ODGI}}$, in Fig. \ref{wasp_wing}(b)-(c) respectively, the total region is divided into 9 sub-regions and the protocol applied to each of them.  At this digitization rate the presence of higher electronic noise lowers the improvement offered by ODGI, by the way few more details can be appreciated in Fig. \ref{wasp_wing}(c) rather than in Fig. \ref{wasp_wing}(b)  (see for example the increased visibility of the edge of the wing in the left side).
In Fig. \ref{detail}(b) the reconstruction of the green bug wing is shown (Fig. \ref{detail}(a) is the direct image). In this case we acquired $4 \cdot 10^4$ frames at 100 kHz. %At this acquisition rate the electronic noise is lower and the improvement of the ODGI technique over the conventional protocol is more significant.
%Being the efficiency of our set-up $\eta \sim 80\%$ the expected difference between $S_{ODGI}$ and $S_{DGI}$ is not enough to be appreciated at sight, therefore only the ODGI reconstructions are reported.
In particular, in Fig. \ref{detail}(c)-(d)  we report the reconstruction of two different details. In Fig. \ref{detail}(c), the edge of the wing is more defined when using $S_{ODGI}$ with respect to the conventional protocol. In Fig. \ref{detail}(d) the $S_{GI}$ and $S_{ODGI}$ reconstructions of a second detail are reported for different number of frames: it can be appreciated that using the ODGI protocol the spot emerges from the noise for a lower number of frames. In particular, note that the detail for 40.000 frames using $S_{GI}$ is comparable with the one obtained for 20.000 by $S_{ODGI}$. This result is in agreement with the prediction of our model for these specific conditions. This means that the photon dose can be reduced of  a factor 2 while providing the same information,  an important improvement when considering delicate samples (e.g. for X-ray ghost imaging).  %(i.e. for $t_-=0.4$ and $\epsilon = 0.1$ it is predicted an advantage of $SNR_{ODGI}/SNR_{GI} \sim 1.5$, which correspond to $1.5^2$ frames equivalent).

\section*{Conclusion}
%In this work we extended the differential ghost imaging protocol [1], to low brightness sources and to the quan-tum case, both theoretically and experimentally.
In this work, we extended the differential ghost imaging protocol \cite{DGI} in the unexplored regime of low brightness sources and quantum correlation. The attention toward the DGI is justified since it offers significant SNR improvements over the conventional GI, in particular in presence of small or highly transparent objects. However, this advantage has been demonstrated in previous works within the implicit assumption of classical intense thermal beams. In this paper in particular we point to the investigation of the low-intensity regime, including quantum resources, which is of great interest since there are practical situations where it is worth to keep low the photon dose, e.g. biological, delicate or photo-sensitive samples. For example, this represents an issue in X-ray imaging and spectroscopy \cite{xray1, xray2}. We note that in this contest, improving the SNR performance at fixed exposure time, i.e. without increasing the photon dose, is of utmost importance.

A theoretical model in terms of experimental quantities has been developed, both for thermal and SPDC light, for any value of the source brightness. Also experimental imperfections such as losses and electronic noise in the detector has been considered. It comes out that DGI performances in the regime of small number of photons per spatio temporal mode ($n_2/M \ll \eta$) are highly affected from experimental imperfections. For example, the DGI advantage in terms of SNR is almost washed away even for relatively high efficiency as the one in our set-up, $\eta=0.8$ and the model shows that for lower efficiencies, namely $\eta<0.5$, DGI is worse than GI. 

Therefore, inspired by what done in the absorption estimation framework in \cite{moreau}, we propose an optimized protocol (ODGI) able to partially compensate for detrimental effects of experimental imperfections such as channel efficiencies and electronic noise, but requiring their estimation.
In the high brightness regime ODGI is always equivalent to the original DGI protocol.  In the opposite, low brightness, case it coincides with DGI when $\eta=1$ while in the realistic condition of $\eta<1$ it  performs always better than both DGI and GI.

The theoretical model has been experimentally validated  in the low brightness regime using quantum correlated beams produced by SPDC. The number of photons collected  per pixel was $n_2 \sim 10^3$: this regime allows to reconstruct the images by performing intensity correlations, without the need of any time-coincidence scheme. Finally, in view of possible real applications, the optimized protocol has been successfully employed in the reconstruction of a complex biological object, demonstrating that a reduction of  the photon dose is possible  while maintaining the same SNR of the conventional GI protocol.

\section*{Acknowledgments}
The authors acknowledge the European Union’s Horizon 2020 and the EMPIR Participating States in the context of the project 17FUN01 BeCOMe for financial support. The authors also thank Matteo Fretto which realized the absorbing samples used.

\section*{Author Contributions}
IRB and EL conceived the idea and developed the theoretical model of this work, that was discussed and designed with inputs by all authors. EL, AM, AA and OS realized the experimental setup and collected the data in INRIM quantum optics labs (coordinated by MG). All authors discussed the results and contributed to the writing of the paper.

\section*{Conflicts of interest}
There are no conflicts of interest.

\end{document}